\documentclass[twocolumn,prb]{revtex4}

\usepackage[dvips]{graphicx}

\usepackage{amsmath}
\usepackage{hyperref}
\usepackage{psfrag}

\newcommand{\p}{\partial}

\begin{document}

\title{The Minkowski metric in non-inertial observer radar coordinates}\thanks{The published version is: \href{http://dx.doi.org/10.1119/1.2060716}{E. Minguzzi, Am. J. Phys. {\bf 73},  1117-1121 (2005)}.}

\author{E. Minguzzi}
\email{minguzzi@usal.es}
\affiliation{Departamento de Matem\'aticas,
Universidad de Salamanca, Plaza de la Merced 1-4, E-37008
Salamanca, Spain and INFN, Piazza dei Caprettari 70, I-00186
Roma, Italy}


\begin{abstract}
We give a closed expression for the Minkowski ($1+1$)-dimensional
metric in the radar coordinates of an arbitrary non-inertial
observer $O$ in terms of $O$'s proper acceleration. Knowledge of
the metric allows the non-inertial observer to perform experiments
in spacetime without making reference to inertial frames. To
clarify the relation between inertial and non-inertial observers
the coordinate transformation between radar and inertial
coordinates, also is given. We show that every conformally flat
coordinate system can be regarded as the radar coordinate system
of a suitable observer for a suitable parametrization of the
observer worldline.  Therefore, the coordinate transformation
between arbitrarily moving observers is a conformal transformation
and conformally invariant ($1+1$)-dimensional theories lead to the
same physics for all observers, independently of their relative
motion.
\end{abstract}

\maketitle

\section{Introduction}

Although the structure of special relativity is best explained
using inertial frames, it is natural to ask how Minkowski
spacetime would appear to a non-inertial observer. The motion of a
non-inertial observer is often studied in an inertial reference
frame. However, it is more convenient to construct coordinates for
the non-inertial observer and to express the metric in these
coordinates. In this way the non-inertial observer does not need
to communicate directly with inertial observers to interpret the
results of her own experiments. Two observations will guide the
mathematical derivations of this work. (a) The definition of the
coordinates must be operationally clear. In ($1+1$)-dimensions
there are many possible choices, but only radar coordinates are
clear operationally. Other well known coordinates, such as Fermi
coordinates,\cite{fermi22,misner73,marzlin94,nesterov99} are
mathematically convenient but turn out to be difficult to
interpret operationally. (b) The motion of the non-inertial
observer influences the expression of the metric in the
non-inertial coordinates. It is convenient to express the motion
of the non-inertial observer by means of her proper acceleration
because this quantity is directly observable and makes no
reference to external inertial frames. Following this strategy the
non-inertial observer metric is expected to depend on the proper
acceleration history of the non-inertial observer.

We shall be particularly concerned with what can be actually
measured by the non-inertial observer, and our construction will depend
on an observable definition of the coordinates (radar coordinates) and
on an observable and intrinsic description of the non-inertial
observer motion (the proper acceleration).

 We consider a non-inertial
observer $O$ in unidirectional, but otherwise arbitrary, motion in
the Minkowski two-dimensional spacetime of metric signature
$(+-)$. The Minkowski metric is given by $d s^{2}= d t^{2}-d
x^{2}$, where $x$ and $t$ are coordinates of an inertial frame $K$
and where units have been chosen such that $c=1$. Let
$\gamma(\tau)$ be the arbitrary timelike worldline of the
non-inertial observer, $\tau$ the proper time parametrization, and
$u^{\mu}=d x^{\mu}/d \tau$ with $\mu=0,1$ the covariant velocity.

As mentioned, the non-inertial observer can construct
coordinates using different methods. The transformation between
inertial and non-inertial coordinates should reduce to a Lorentz
transformation if the observer has vanishing acceleration.
However, this requirement is too weak to determine a unique
solution. There are at least three possibilities that have been
considered: radar
coordinates,\cite{landau62,marzke64,pauri00,dolby01} Fermi
coordinates\cite{fermi22,misner73,marzlin94,nesterov99} and
M\o{}ller coordinates.\cite{brehme62,sears68}

Even more possibilities arise if the observer is not regarded as isolated,
but as part of an extended reference frame\cite{footnote} where no worldline
has a privileged role.\cite{manoff01,minguzzi03} In this case the coordinate
system also would depend on the rigidity assumptions that constrain the
frame.\cite{rosen47,romain63,romain64b,bel95}

Here we shall focus on radar coordinates. They are commonly used
in the study of accelerating\cite{pauri00,dolby01} and rotating
systems in special relativity.\cite{davies76,ashworth76} These coordinates
are not related to measurements performed with standard rods and clocks.
They are just convenient labels of spacetime events. However, in the
neighborhood of the observer worldline, because they approximate
the coordinates of a comoving inertial frame, they represent
measurements performed by rods and clocks. For example, the radar
distance becomes the distance that would be measured using
standard rods.\cite{landau62}

\section{Radar coordinates and proper acceleration}\label{radar}

Radar coordinates are defined as follows. Let $E$ be a
spacetime event and consider the light beam $L_2$ emitted at
$E$. It reaches $O$'s worldline at the proper time $\tau_2(E)$.
Consider also the light beam $L_1$ that was emitted at a suitable
proper time $\tau_1(E)$ and reaches $E$. By using the radar procedure, the
observer $O$ assigns to $E$ a time label given by
$(\tau_2(E)+\tau_1(E))/2$. Note that if $E=\gamma(\tau)$,
then the time label assigned to $E$ is just $\tau$. For this
reason the time label will be denoted by the same symbol
\begin{equation}
\tau(E)=\frac{\tau_2(E)+\tau_1(E)}{2}.
\end{equation}
Next $O$ assigns to the event a radar distance given by
\begin{equation}
d(E)=\frac{\tau_2(E)-\tau_1(E)}{2}.
\end{equation}
It is more convenient to define a spatial coordinate
$\chi$ rather than a distance, and we define $\chi=\pm d$
with the plus sign if $E$ stays to the right of the curve
$\gamma(\tau)$ or the minus sign if $E$ stays to the left.
\begin{figure}[!ht] \label{radar2}
\centering \psfrag{A}{$\tau^{+}(A)$} \psfrag{B}{$\tau^{-}(A)$}
\psfrag{C}{$\tau^{+}(B)$}\psfrag{D}{$B$}
\psfrag{E}{$\!\!\!\!\!\!\!\!\!\tau^{-}(B)$} \psfrag{F}{$A$}
\psfrag{H}{$\gamma(\tau)$}
\includegraphics[width=4cm]{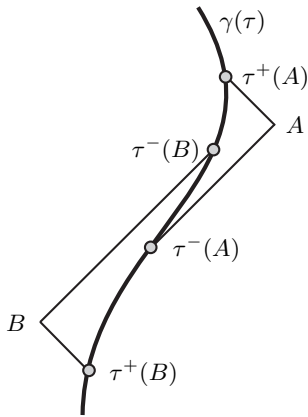}
\caption{Definition of $\tau^{+}$ and $\tau^{-}$.}
\end{figure}
It also is convenient to introduce the quantities $\tau^{+}(E)$
and $\tau^{-}(E)$ (see Fig.~\ref{radar2}). These quantities
coincide respectively with $\tau_{2}(E)$ and $\tau_{1}(E)$ if $E$
stays to the right of the curve $\gamma(\tau)$, and in the
opposite case, they coincide respectively with $\tau_{1}(E)$ and
$\tau_{2}(E)$. We can say that $\tau^{+}$ determines the
intersection event between $\gamma$ and a light beam moving from
the right to the left, and $\tau^{-}$ determines the intersection
event between $\gamma$ and a light beam moving from the left to
the right. With these definitions we have the radar coordinates
\begin{eqnarray}
\tau(E)&=&\frac{\tau^{+}(E)+\tau^{-}(E)}{2} ,\\
\chi(E)&=&\frac{\tau^{+}(E)-\tau^{-}(E)}{2} ,
\end{eqnarray}
where $\tau^{+}(E)$ and $\tau^{-}(E)$ are directly observable
quantities. The radar coordinates are defined in the intersection
between the causal future and the causal past of the curve
$\gamma$. Outside this region, $\tau_2$, or $\tau_1$, or both,
are not defined. A typical example is the constant acceleration
case. Only the events in the Rindler wedge, $-x<t<x$, admit two
light beams passing through them that intersect the curve
$\gamma$, thus defining $\tau_2$ and $\tau_1$.

If we introduce the coordinates $x^{\pm}=t\pm x$, then the
Minkowski metric takes the form
\begin{equation}
d s^{2}= d x^{+} d x^{-} .
\end{equation}
Note that the events with a given value of $\tau^{\pm}$ lie in a plane
of equation $x^{\pm}=\textrm{constant}.$ This fact will be central in
what follows and justifies the introduction of $\tau^{+}$ and
$\tau^{-}$ in place of $\tau_2$ and $\tau_1$.

Finally, we introduce the acceleration of the non-inertial
observer given by $a^{\mu}=\nabla_u u^{\mu}$, where $\nabla$ is
the Levi-Civita connection of the Minkowski metric $d s^2=g_{\mu
\nu} d x^{\mu} d x^{\nu}$, that is, if $v$ is a vector field,
$(\nabla_{u} v)^\alpha=v^{\alpha}_{,\, \mu} u^{\mu}
+\Gamma^{\alpha}_{\nu \mu} v^{\nu} u^{\mu}$, where $
\Gamma^{\alpha}_{\nu \mu}=\frac{1}{2} g^{\alpha \beta}(g_{\beta
\mu,\nu} +g_{\beta \nu,\mu}-g_{\mu\nu, \beta})$ are the
Christoffel coefficients. In inertial frame coordinates $g_{\mu
\nu}=\eta_{\mu \nu}$ and $\Gamma^{\alpha}_{\mu \nu}=0$. Because
$u \cdot u=1$, we have $a \cdot u=0$, and hence $a$ is spacelike.
Let $e_1$ be the normalized ($e_1 \cdot e_1 =-1$) spacelike vector
orthogonal to $u$ and oriented toward increasing $x$ ($e_1[x] > 0$).
Because $a \cdot u=0$, we have
\begin{equation} \label{acce}
a^{\mu}(\tau)=a(\tau) e_1^{\mu}(\tau),
\end{equation}
where $a$ is the acceleration of $O$ with respect to the local
inertial frame. In other words, $-a(\tau)$ is the apparent
acceleration of free particles with respect to the accelerated
observer.

\section{The Minkowski metric in radar and Fermi coordinates}\label{mmetric}
Our goal is to prove the the following result:
In the radar coordinates $(\tau, \chi)$, the Minkowski metric takes
the form
\begin{equation}
\label{f}
d s^{2}=e^{\int^{\tau+\chi}_{\tau-\chi} a(\tau') d \tau'} (d
\tau^{2}-d \chi^{2}) .
\end{equation}
This result generalizes to arbitrary motion the constant
acceleration case considered by Lass\cite{lass63} for which he finds, given
certain assumptions, that
\begin{equation}
d s^{2}=e^{ 2a\chi} (d \tau^{2}-d \chi^{2}) .
\end{equation}
The fact that the coordinates $\tau$ and $\chi$ are radar
coordinates of the uniformly accelerated observer was pointed out
later.\cite{desloge87} The constant acceleration case is also
discussed in
Refs.~\onlinecite{romain63,romain64,marsh65,tinbrook97}. It is
remarkable that the problem can be solved without restricting
assumptions. Note that at time $\tau$ the non-inertial observer
knows the spacetime coordinates $(\tau,\chi)$ of the events in the
causal past of $\gamma(\tau)$. At the same time the observer can
immediately calculate the metric in its causal past from the
knowledge of the acceleration $a$ as measured by a comoving
accelerometer. It also is interesting that the metric at the event
$E$ is determined by the acceleration of the observer in events
that are spacelike separated from $E$. Indeed, the integral that
appears in Eq.~(\ref{f}) is between $\tau^{-}(E)$ and
$\tau^{+}(E)$. From Eq.~(\ref{f}) we see that the radar
coordinates are conformally related with the inertial coordinates.
Thus the speed of light in radar coordinates is equal to one in
both directions,\cite{romain63} indeed $d s^{2}=0$ implies $\vert
d \chi/d \tau \vert=1$.

The proof of \eqref{f} goes as follows. Because the level set of the function
$\tau^{\pm}$ coincides with that of the function $x^{\pm}$, there exist
functions $g^{+}$ and $g^{-}$ such that
\begin{subequations}
\label{8}
\begin{align}
x^{+}&=g^{+}(\tau^{+}) , \label{1}\\
x^{-}&=g^{-}(\tau^{-}). \label{2}
\end{align}
\end{subequations}
We therefore have
\begin{subequations}
\begin{align}
d s^{2}&= {g^{+}}'(\tau^{+}) {g^{-}}'(\tau^{-}) d \tau^{+}
d \tau^{-} \\
&={g^{+}}'(\tau^{+}) {g^{-}}'(\tau^{-})(d \tau^{2}-d
\chi^{2}).
\end{align}
\end{subequations}
The worldline $\gamma$ satisfies $\chi=0$, and along this
worldline, $d s=d \tau$ and $\tau^{+}=\tau^{-}=\tau$; thus
${g^{+}}'(\tau) {g^{-}}'(\tau)=1$. This argument, which also can
be found in Jones,\cite{jones61} allows us to conclude that there
is a function $G(\tau)$ such that
\begin{equation}
\label{3}
g^{\pm}(\tau)=\!\int_{0}^{\tau} e^{\pm G(\tau')} d \tau' +C^{\pm} ,
\end{equation}
where $C^{\pm}$ is an integration constant. The metric takes the
form
\begin{equation}
\label{oh}
d s^{2}=e^{G(\tau+\chi)-G(\tau-\chi)} (d \tau^{2}-d
\chi^{2}).
\end{equation}
This metric shows that $e_{1}=\p/\p \chi$ along the worldline
$\chi=0$.

Now we make a little digression and consider a different
problem. Let $g'_{\mu \nu}=\Omega^{2} g_{\mu \nu}$ be two
conformally related metrics. The geodesics of $g'_{\mu \nu}$ are
extremals of the length functional
\begin{equation}
I[\gamma]=\!\int_{\gamma} \sqrt{g'_{\mu \nu}\frac{d x^{\mu}}{d
\lambda} \frac{d x^{\nu}}{d \lambda} }\, d
\lambda=\!\int_{\gamma}\Omega \sqrt{g_{\mu \nu}\frac{d
x^{\mu}}{d \lambda}\frac{d x^{\nu}}{d \lambda} }\, d
\lambda .
\end{equation}
Let $u^{\mu}=d x^{\mu} / d s$ with $d s^{2}=g_{\mu \nu} d x^{\mu}
d x^{\nu}$. If we take the variational derivative of $I[\gamma]$,
we obtain
\begin{equation}
g_{\mu \nu}\nabla_{u} u^{\nu}= (\delta^{\alpha}_{\mu}-u^{\alpha}
u_{\mu}) \p_{\alpha} \ln \Omega ,
\end{equation}
where $\nabla$ is the Levi-Civita connection of $g_{\mu \nu}$.

We return to our original problem, and let $g=d t^{2}-d
x^{2}$ be the Minkowski metric, $g'=d \tau^{2}-d \chi^{2}$, and
$\Omega^{-2}=e^{G(\tau+\chi)-G(\tau-\chi)}$. The worldline
$\chi=0$ is clearly a geodesic for the metric $g'$, and therefore
if $u^{\mu}$ and $ a^{\mu}$ are the covariant velocity and
acceleration of $O$, we have
\begin{equation}
\label{prod}
a_{\mu}=-\frac{1}{2}(\delta^{\alpha}_{\mu}-u^{\alpha} u_{\mu})
\big[\p_{\alpha}
G(\tau+\chi)\vert_{\chi=0}-\p_{\alpha}G(\tau-\chi)\vert_{\chi=0}\big].
\end{equation}
The product of \eqref{prod} with $e_{1}^{\mu}$ gives
\begin{equation}
a(\tau)=\frac{1}{2} [\p_{\chi}
G(\tau+\chi)\vert_{\chi=0}-\p_{\chi}G(\tau-\chi)\vert_{\chi=0}]
= G'(\tau),
\end{equation}
from which it follows that
\begin{equation}
\label{that}
G(\tau)=\!\int^{\tau}_{0}a(\tau')d \tau'+G(0) .
\end{equation}
If we substitute Eq.~\eqref{that} into Eq.~(\ref{oh}), we obtain
Eq.~(\ref{f}).

The coordinate transformation between $(\tau,\chi)$ and $(t,x)$
can be found from Eqs.~(\ref{8}) and (\ref{3})
\begin{eqnarray}
t&=& t(0)+\frac{1}{2}\sqrt{\frac{1+v(0)}{1-v(0)}}\!\int_{0}^{\tau
+\chi}
e^{\int^{\tau'}_{0}a(\tau'')d \tau''}d \tau' \nonumber\\
&&{}+\frac{1}{2}\sqrt{\frac{1-v(0)}{1+v(0)}}\!\int_{0}^{\tau -\chi}
e^{-\int^{\tau'}_{0}a(\tau'')d \tau''} d \tau' ,
\label{tr1}\\
x&=&x(0)+\frac{1}{2}\sqrt{\frac{1+v(0)}{1-v(0)}}\!\int_{0}^{\tau
+\chi} e^{\int^{\tau'}_{0}a(\tau'')d \tau''}d
\tau' \nonumber\\
&&{}-\frac{1}{2}\sqrt{\frac{1-v(0)}{1+v(0)}}\!\int_{0}^{\tau -\chi}
e^{-\int^{\tau'}_{0}a(\tau'')d \tau''} d \tau' , \label{tr2}
\end{eqnarray}
where $v(0)$ and and $(t(0),x(0))$ are the velocity and position
of $O$ with respect to $K$ at $\tau=0$. For $a=0$ this
transformation becomes a Poincar\'e transformation between an
inertial frame moving at speed $v(0)$ and $K$.

It is interesting to compare these results with the analogous
results obtained for the Fermi coordinates. The accelerated
observer sets up Fermi coordinates $(\tau_f,\chi_f)$ if two
conditions are fulfilled: (1) the simultaneity slices,
$\tau_f=\mbox{constant}$, are lines perpendicular to the observer
worldline, and (2) given the event $E$, $\chi_f(E)$ is, up to a
sign, the length of the geodesic segment that lies in the
simultaneity slice and that connects $E$ with the observer's
worldline. In the $1+1$ case, the coordinate transformation from
Fermi to inertial coordinates was derived in
Ref.~\onlinecite{born58} and studied in
Ref.~\onlinecite{crampin59}. The main result of
Refs.~\onlinecite{born58} and~\onlinecite{crampin59} is quite
elegant. We summarize it in the following. For a proof we refer
the reader to the original work\cite{born58,crampin59} or to the
problems in Sec.~\ref{problems} where a more direct derivation is
proposed.

Let $(T(\tau),X(\tau))$ be the trajectory of the non-inertial
observer $O$ with respect to an inertial frame of coordinates
$(t,x)$, where $\tau$ is $O$'s proper time, and let
$(\tau_f,\chi_f)$ be Fermi coordinates set up by $O$. Then
\begin{align}
t&=T(\tau_{f})+ \chi_{f} \dot X(\tau_f), \label{trf1}\\
x&=X(\tau_{f})+\chi_{f} \dot T(\tau_f), \label{trf2}
\end{align}
and the metric takes the form
\begin{equation}
d s^2=[1+a(\tau_{f})\chi_f]^2 d \tau_f^2-d \chi_f^2,
\label{metr2}
\end{equation}
where $a(\tau)$ is the acceleration with respect to the local
inertial frame. For a constant acceleration $a$, this metric is well known.
It is remarkable that it changes only minimally, through the dependence
of $a$ on the proper time, when $a$ is allowed to change.

\section{An inverse result}\label{inverse}
The radar coordinates of $O$ are obtained from the proper time
measurements $\tau^{+}$ and $\tau^{-}$. Suppose that $O$
parametrizes the worldline $\gamma$ using a new parameter
$\mathcal{T}=f(\tau)$ with $f'>0$. Moreover, suppose that $O$
still makes use of the radar method in order to extend a
coordinate system over spacetime. In place of $\tau^{\pm}$, the
observer measures $\mathcal{T}^{\pm}=f(\tau^{\pm})$ and assigns to the
generic event $E$ with coordinates $(\tau,\chi)$ the coordinates
\begin{align}
\mathcal{T}&=\frac{f(\tau+\chi)+f(\tau-\chi)}{2} , \label{1a}\\
\mathcal{X}&=\frac{f(\tau+\chi)-f(\tau-\chi)}{2} . \label{2a}
\end{align}
Transformations of this form are called {\em worldline
reparametrizations} independently of the physical interpretation
of the coordinates. Given a worldline $\gamma$, the worldline
reparametrizations form a group of coordinate transformations; in this
group there is a particular choice of coordinates that corresponds to the
proper time parametrization that ultimately is used in practice. In what
follows all the coordinate systems in the group will be considered as radar
coordinates for the worldline $\gamma$. We shall say that
$(\mathcal{T},\mathcal{X})$ are radar coordinates of the
parametrized curve $\gamma(\mathcal{T})$, while $(\tau,\chi)$ are
radar coordinates of the parametrized curve $\gamma(\tau)$ (proper
radar coordinates).

We found in Sec.~\ref{mmetric} that the radar coordinates
are conformally related with the inertial coordinates. Now we
give a result that may be thought of as the inverse of this result.
We recall that $(\mathcal{T},\mathcal{X})$ are conformally flat
coordinates if the metric (in our case the Minkowski metric) takes
the form $d s^{2}=\Omega(\mathcal{T},\mathcal{X})^{2}(d
\mathcal{T}^{2}-d \mathcal{X}^{2})$ for a suitable positive
function $\Omega$. Let $R$ be the coordinate transformation from
the inertial coordinates $(t,x)$ to the conformally flat
coordinates $(\mathcal{T},\mathcal{X})$, that is,
$(\mathcal{T},\mathcal{X})=R(t,x)$, and define the coordinates
$(\mathcal{T}',\mathcal{X}')=R \circ P(t,x)$, where $P$ is a
Poincar\`e transformation. Then the spacetime metric in the primed
coordinates has the same functional dependence as the unprimed
coordinates, $d s^{2}=\Omega(\mathcal{T}',\mathcal{X}')^{2} (d
{\mathcal{T}'}^2-d {\mathcal{X}'}^{2})$. We stress that
$(\mathcal{T}$, $\mathcal{X})$, and $(\mathcal{T}',\mathcal{X}')$
are distinct conformally flat coordinates.

Note that if $(\mathcal{T},\mathcal{X})$ is a conformally flat
coordinate system, then $\p_{\mathcal{T}}$ is timelike and either
past-pointing or future-pointing. Analogously $\p_{\mathcal{X}}$
is spacelike and either left-pointing or right-pointing. We
restrict our attention to a conformally flat coordinate system for
which $\p_{\mathcal{T}}$ is future-pointing and
$\p_{\mathcal{X}}$ is right-pointing. Otherwise, we have only to
make the replacement $\mathcal{T} \to -\mathcal{T}$ or/and
$\mathcal{X} \to -\mathcal{X}$. We shall call them {\em oriented
conformally flat coordinate systems}.

Our goal is to prove the following statement: If
$(\mathcal{T},\mathcal{X})$ are oriented conformally flat
coordinates in $1+1$ Minkowski spacetime, then
$(\mathcal{T},\mathcal{X})$ are radar coordinates of the
parametrized curve $\gamma(\mathcal{T})$ of the equation
$\mathcal{X}=0$. An important consequence of this result is that
up to worldline reparametrizations, there is a one-to-one
correspondence between oriented conformally flat coordinate
systems in $1+1$ Minkowski spacetime and observers (timelike
worldlines).

The proof goes as follows. Define
$\mathcal{T}^{\pm}=\mathcal{T}\pm\mathcal{X}$, the level sets
$\mathcal{T}^{\pm}=\textrm{constant}$ are null, and hence
$\mathcal{T}^{\pm}$ is a invertible function of $x^{\pm}$. (We use the
fact that $\p_\mathcal{T}$ is future-pointing and $\p_\mathcal{X}$
is right-pointing.) If we use this functional dependence in the
metric, we obtain $d x^{+} d x^{-}=\Omega^{2}d \mathcal{T}^{+} d
\mathcal{T}^{-}$, and it follows that
$\Omega^{2}=\exp[H^{+}(\mathcal{T}^{+})+H^{-}(\mathcal{T}^{-})]$
for suitable functions $H^{-}$ and $H^{+}$. The same result could
have been derived by noting that the Ricci scalar in two spacetime
dimensions reads,\cite{wald84}
$R=-\Omega^{-2}(\p^{2}_{\mathcal{T}}-\p^{2}_{\mathcal{X}})\ln
\Omega^{2}$; because we are assuming a flat metric, $R=0$. We
define the functions
\begin{align}
F&=\frac{H^{+}+H^{-}}{2} ,\\
\tilde{G}&=\frac{H^{+}-H^{-}}{2} .
\end{align}
Then
\begin{equation}
d
s^{2}=e^{\tilde{G}(\mathcal{T}^{+})-\tilde{G}(\mathcal{T}^{-})}
e^{F(\mathcal{T}^{+})+F(\mathcal{T}^{-})} d \mathcal{T}^{+} d
\mathcal{T}^{-} .
\end{equation}
Next, introduce the coordinates $(\tau,\chi)$ and the
corresponding $\tau^{\pm}$ through the worldline reparametrization,
Eqs.~(\ref{1a}) and (\ref{2a}), or $\tau^{\pm}=f^{-1}(\mathcal{T}^{\pm})$,
$f^{-1}(f(y))=y$ with $f^{-1}(\mathcal{T})=\!\int^{\mathcal{T}}_{0}
e^{F(\mathcal{T}')} d \mathcal{T}'$. In addition, we define
$G(\tau^{\pm})=\tilde{G}(f(\tau^{\pm}))$ and obtain
\begin{equation}
\label{formula}
d s^{2}=e^{G(\tau^{+})-G(\tau^{-})} d \tau^{+}d \tau^{-} .
\end{equation}
Let $\gamma$ be the timelike curve $\mathcal{X}=0$. The
Eq.~\eqref{formula} shows that $\tau$ is the proper time
parametrization of $\gamma$. Indeed,   the spacetime functions
$\mathcal{T}^{\pm}$ coincide on $\gamma$ with the curve
parametrization
$\mathcal{T}^{\pm}\vert_{\gamma}=\mathcal{T}\vert_{\gamma}$, and hence
$\tau^{+}\vert_{\gamma}=\tau^{-}\vert_{\gamma}=\tau$ and from
Eq.~\eqref{formula}, $d s= d \tau$. According to the results of
Sec.~\ref{mmetric}, Eq.~\eqref{formula} shows that the
acceleration of $\gamma$ is $a(\tau)=G'(\tau)$. The coordinates
$(\tau,\chi)$ are therefore the proper radar coordinates of
$\gamma$, and hence $(\mathcal{T},\mathcal{X})$ are radar
coordinates with respect to the worldline parametrization
$\gamma(\mathcal{T})$.

\section{Summary}
We have given the transformation between the radar coordinates of
an arbitrarily accelerating observer and the radar coordinates of
an inertial observer. A closed expression for the spacetime metric
in the former coordinates in terms of the acceleration of the
non-inertial observer also has been given. Then we showed that
every oriented conformally flat coordinate system represents, up
to a worldline reparametrization, an observer. Because every
observer is represented by an oriented conformally flat coordinate
system, the coordinate transformation between two arbitrary
observers is a conformal transformation. Therefore, a
$(1+1)$-dimensional theory invariant under conformal transformations
would give rise to the same physics for all the
observers independently of their relative motion.
This symmetry should be broken by a
Lagrangian that defines a realistic physical theory. Otherwise, it
would be impossible to measure the acceleration with an
accelerometer.

These considerations are related to analogous conclusions that
have been reached in higher dimensional cases. Because the
conformal group has only fifteen generators for more than two
dimensions, it can represent only the coordinate transformations
between uniformly accelerated observers.\cite{hill45,hill47} (This
case should be compared with the two-dimensional case considered
here where the conformal group has an infinite number of
generators.) Because Maxwell's equations are invariant under
conformal transformations, they determine a physics that would
appear to be the same in uniformly accelerating and inertial
frames. However, the mass term in the Lorentz force equation
ultimately spoils the symmetry, leading to the possibility of
determining the acceleration of a frame by looking at the motion
of charged particles.\cite{fulton62,fulton62b}

\section{Suggested PROBLEMS}\label{problems}

\noindent Problem 1. Verify that Eqs.~(\ref{tr1}) and (\ref{tr2})
reduce to the
Lorentz transformation for $a(\tau)=0$, $t(0)=x(0)=0$.

\noindent \smallskip Problem 2. Verify that the coordinates ($\tau_f$,
$\chi_f$), which satisfy Eqs.~(\ref{trf1}) and (\ref{trf2}), are
Fermi coordinates by proving that (a) the simultaneity slices,
$\tau_f=\mbox{constant}$, are lines orthogonal to the observer
worldline. (Use the fact that the observer's covariant velocity
$u=(\dot{T},\dot{X})$ is orthogonal to the normalized spacelike
vector $e_1=(\dot{X},\dot{T})$ that enters in the transformation
law in vectorial form.) (b) The distance of the observer
worldline $(T(\tau),X(\tau))$ from the event $(t,x)$ along the
simultaneity slice is given by $\chi_f$ (because $\tau$ is a
proper time $\dot{T}^2-\dot{X}^2=1$). Recall from Eq.~\eqref{acce} that
$a^{\mu}=a(\tau)e_1^{\mu}$, where $a(\tau)$ is the acceleration of $O$ with
respect to the local inertial frame. Prove that
$\dot{T}(\tau)=a(\tau)\dot{X}(\tau)$ and
$\dot{X}(\tau)=a(\tau)\dot{T}(\tau)$ and use these relations to
derive Eq.~(\ref{metr2}).

\begin{acknowledgments}
The author is supported by INFN, grant \# 9503/02.
\end{acknowledgments}


\begin{thebibliography}{27}
\bibitem{footnote}{An
extended frame is defined by a congruence of timelike curves, while
an observer in 1+1 Minkowski spacetime is defined by a worldline.}
\end{thebibliography}
\end{document}